\documentclass[prl,twocolumn,letterpaper,superscriptaddress]{revtex4-2}
\usepackage{graphicx,amsmath,amssymb,amsfonts,latexsym,color,dcolumn,bm,epsfig,subfigure}
\usepackage[plainpages=false,hyperfootnotes=false,colorlinks=false]{hyperref}
\usepackage[normalem]{ulem}
\usepackage{dsfont}
\usepackage{tikz}
\usepackage{comment}
\usepackage{mathrsfs}
\usepackage{amsbsy}
\usepackage{mhchem}
\renewcommand{\imath}[0]{\mathrm{i}}

\usepackage{orcidlink}

\begin{document}

\title{Nonequilibrium Casimir-Polder Force: Magnus-like Effect}

\author{Maria Vittoria Gurrieri\orcidlink{0000-0002-1654-3339}}
\affiliation{Humboldt-Universit\"at zu Berlin, Institut f\"ur  Physik, AG Theoretische Optik and Photonik, 12489 Berlin, Germany}

\author{Kurt Busch\orcidlink{0000-0003-0076-8522}}
\affiliation{Humboldt-Universit\"at zu Berlin, Institut f\"ur  Physik, AG Theoretische Optik and Photonik, 12489 Berlin, Germany}
\affiliation{Ma$x$-Born-Institut, 12489 Berlin, Germany}

\author{Francesco Intravaia\orcidlink{0000-0001-7993-4698}}
\affiliation{Humboldt-Universit\"at zu Berlin, Institut f\"ur  Physik, AG Theoretische Optik and Photonik, 12489 Berlin, Germany}

\newcommand{\fran}[1]{{\color{blue}#1}}

\begin{abstract} 
The motion of a particle in vacuum near macroscopic bodies gives rise to a Magnus-like contribution to the nonequilibrium Casimir-Polder force. This effect originates from the interplay between particle dynamics and material-modified electromagnetic quantum fluctuations, inducing in the particle a direction-dependent angular momentum coupled to the electromagnetic field spin. The resulting drift force is proportional to the cross product of the particle's angular and translational velocities, revealing a rotational transport component in the nonequilibrium Casimir-Polder interaction. Our results establish a striking connection between quantum fluctuations-induced forces and the classical Magnus effect in fluid dynamics.
\end{abstract}

\maketitle


A spinning object moving through a fluid experiences a transverse force proportional to the cross product of its angular and linear velocity~\cite{Munson09}, a phenomenon known as the Magnus effect~\cite{Magnus53}. This force plays a central role in classical fluid dynamics, with applications ranging from aviation to ship propulsion~\cite{De-Marco16, Bordogna20,Seifert12a}. The transverse force associated with the Magnus effect has also been investigated in more exotic contexts, such as superfluidity, superconductors, and gravitational interactions~\cite{Donnelly69,Sonin97,Bellizotti-Souza22,Costa18,Wang24}, revealing the broad relevance of rotationally induced transport phenomena. 

At the nanoscale the emergence and persistence of a Magnus-type force is nontrivial~\cite{Changfu03,Solsona20,Cao23}. In this regime, interactions between particles and their environment are often governed by fluctuation-induced forces. A paradigmatic example is the Casimir–Polder interaction, which arises from fluctuations of the quantum electromagnetic field~\cite{Schwinger78} and acts between electrically neutral, non-magnetic particles and nearby material objects~\cite{London30,Casimir48a,Dzyaloshinskii61a}. This interaction is mostly studied under equilibrium conditions and plays an important role in quantum technologies, including atom interferometry~\cite{Cronin09,Hornberger12,Alauze18}, atom-chip platforms~\cite{Henkel99,Reichel11,Schneeweiss12,Keil16,Wongcharoenbhorn21}, and hybrid atom–fiber systems~\cite{Vetsch10,Reitz13}. Equilibrium is, however, an approximation that imposes severe constraints both on the physics of the system and the experimental setup. Consequently, growing attention has turned to out of equilibrium configurations~\cite{Antezza05,Klimchitskaya22a,Milton23,Shen25,Reiche22}, where the relative motion of a particle and a surface gives rise to nonconservative forces. 
In this regime, the interplay of the particle's internal degrees of freedom with the material-modified quantum fluctuations of the electromagnetic field introduces a preferred rest frame, such that the electromagnetic vacuum effectively acquires the character of a quantum fluid and exerts a viscous resistance on the moving particle~\cite{Noether18,Oelschlager22}. 
This behavior originates from the anomalous Doppler mechanism underlying the quantum Cherenkov effect~\cite{Frolov86,Nezlin76,Ginzburg96,Maghrebi13}, whereby the motion of the particle leads to the emission of real excitations, resulting in momentum recoil and kinetic energy loss~\cite{Intravaia15}. When rotational degrees of freedom are included, the spin–momentum locking of light~\cite{Bliokh15,Lodahl17} and the Doppler shift lead to an exchange of angular momentum between the particle and the field. In this process, the particle acquires an angular momentum, thus mimicking a wheel rolling above the surface~\cite{Intravaia19a}. 
\begin{figure}
  \begin{center}
   \includegraphics[width=.9\linewidth]{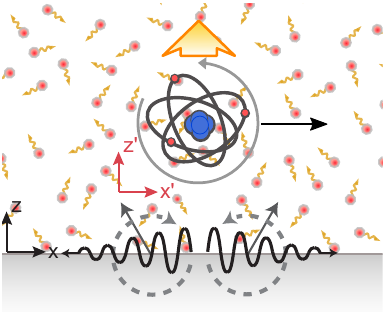}
   \end{center}
   \begin{tikzpicture}[remember picture, overlay]
    \node[rotate=0, font=\large] at (-3.5,4.8) {$T=0$};
        \node[rotate=0, font=\large] at (0.18,6.44) {$\mathbf{F}^{\rm Th}_{\rm a}$};
              \node[rotate=0, font=\large] at (-1.,5.8) {$\boldsymbol{\Omega}$};
               \node[rotate=0, font=\large] at (-1.2,3.7) {\color{red}$ \mathbf{J}_{\overline\chi}$};
                  \node[rotate=0, font=\LARGE] at (2.2,4.3) {$v$};
        \node[rotate=0, font=\normalsize] at (-1.9,2.7) {$\mathbf{E}$};
        \node[rotate=0, font=\normalsize] at (2.1,2.7) {$\mathbf{E}$};
            \node[rotate=0, font=\normalsize] at (-1.6,1.0) {$\boldsymbol{\chi}$};
        \node[rotate=0, font=\normalsize] at (2.,1.0) {$\boldsymbol{\chi}$};
            \node[rotate=0, font=\normalsize] at (-1.6,1.0) {$\boldsymbol{\chi}$};
        \node[rotate=0, font=\normalsize] at (3.,1.7) {$q$};
        \node[rotate=0, font=\normalsize] at (-2.7,1.7) {$-q$};
     \end{tikzpicture}
     \vspace{-0.5cm}
  \caption{Schematic description of the mechanisms underlying the Magnus-like effect, where a polarizable particle moves at a constant height and velocity above a flat surface. The electromagnetic surface excitations carry a spin locked to their direction of propagation~\cite{Bliokh15,Lodahl17}. In the particle’s rest frame, they give rise to a nonzero spin current density. These excitations are also absorbed and re-emitted by the atom and a net angular momentum is absorbed by the particle. For this configuration, the combination of the translational motion and the induced ``rotation", produces a Magnus-like drift force directed along the positive $z$-axis and with sign opposite to the one expected for the Casimir-Polder force.\label{fig:Magnus_force}}
  \vspace{-0.5cm}
\end{figure}
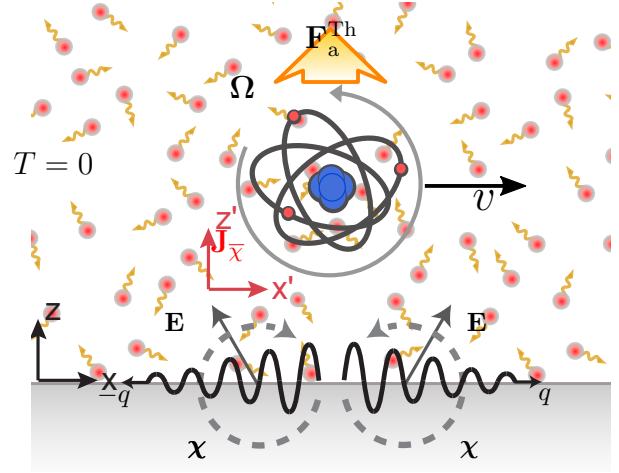
In the presence of both angular and translational motion within an effective fluid, one can ask whether a rotationally induced transverse force, analogous to the classical Magnus effect, can emerge. In this work, we demonstrate that the answer is affirmative and that such a force arises as a distinct contribution to the nonequilibrium Casimir–Polder interaction, in which a term proportional to the cross product of the particle's angular momentum and translational velocity can be isolated. In particular, our results demonstrate that material-modified electromagnetic quantum fluctuations can support Magnus-type phenomena, see~Fig.\eqref{fig:Magnus_force}.

We study a system at zero temperature in which a neutral atom moves at constant, non-relativistic speed $v$ along the translational-invariant direction of a macroscopic ensemble of passive, linear, and reciprocal bodies. The motion is sustained by an external drive that balances the viscous drag force, allowing the system to reach a nonequilibrium steady state (NESS)~\cite{Intravaia19a}. 
The intrinsic magnetic properties of the particle and its surroundings are neglected, and the internal quantum dynamics of the particle are described within the electric dipole approximation by the operator $\mathbf{\hat{d}}(t)$. 
In this regime, the Casimir-Polder force acquires nonequilibrium contributions induced by the particle's motion and its interaction with the Doppler-shifted quantum fluctuations of the electromagnetic field. The force can be decomposed as the sum of three terms~\cite{SuppMat,Reiche26}, 
    $\mathbf{F} = \mathbf{F}^{\rm Ds}+ \mathbf{F}^{\rm Th} +\mathbf{F}^{\rm As}$,
the last two are pure nonequilibrium corrections to the interaction. $\mathbf{F}^{\rm Ds}$ generalizes the equilibrium Casimir-Polder force, which is recovered in the limit of $v \to 0$; $\mathbf{F}^{\rm Th}$ originates from nonequilibrium-induced asymmetric absorption and emission processes involving the particle's internal degrees of freedom; $\mathbf{F}^{\rm As}$ is defined by the geometric properties of the system and vanishes for mirror-symmetric configurations relative to the plane that contains the trajectory of the particle~\cite{Reiche26}. In the following, we disregard $\mathbf{F}^{\rm As}$, since it can vanish even for nonequilibrium configurations, and we focus on $\mathbf{F}^{\rm Th}$. This term can be written as~\cite{SuppMat,Reiche26}
\begin{align}
\label{FCPNEqTH}
\mathbf{F}^{\rm Th}
=\int\limits_{0}^{\infty}\mathrm{d}\omega\int\frac{\mathrm{d}q}{2\pi}~
\mathrm{Tr}\left[\underline{S}^{\sf T}(-\omega^{-}_{q},v) \nabla_{\mathbf{R}_{a}}\underline{G}_{\Re}(q, \mathbf{R}_{a}, \omega)\right],
\end{align}
where $q$ is the component of the wave vector along the direction of motion, $\omega_q^{\pm}=\omega \pm qv$ denotes the Doppler-shifted frequency and $\mathbf{R}_a$ defines the lateral position of the particle within the plane orthogonal to the direction of motion. 
The force involves the trace ($\operatorname{Tr}$) of the dipole's power spectrum tensor $\underline{S}$, describing the quantum statistical properties of the moving particle in the NESS~\footnote{Given that $\underline{S}$ results from an expectation value taken over the NESS, it depends not only on the particle's velocity $v$ but also on its position $\mathbf{R}_{a}$, which we do not shown for simplicity.}, and the Hermitian part of the Fourier transformed electromagnetic Green tensor, i.e. $\underline{G}_{\Re}=[\underline{G}+\underline{G}^{\dagger}]/2$~\cite{Jackson75,Landau80}. Since both $\underline{S}$ and $\underline{G}_{\Re}$ are Hermitian, they admit the decomposition~\cite{Intravaia26}
\begin{equation}
\label{decomposition}
\underline{S}= \underline{K} + \mathbf{\Omega}\cdot \underline{\bm{L}} \quad\text{and}\quad \underline{G}_{\Re}=\underline{\Sigma} + \bm{\chi}\cdot \underline{\bm{L}}~.
\end{equation}
$\underline{K}$ and $\underline{\Sigma}$ are real symmetric matrices, $\left[ L_k \right]_{ij} =- \mathrm{i}\epsilon_{kij}$ are the skew-symmetric matrix generators of the Lie algebra $\bm{\mathfrak{so}(3)}$ and $\mathbf{\Omega}$ as well as $\bm{\chi}$ are real three-dimensional vectors. The trace properties lead to a decomposition of $\mathbf{F}^{\rm Th}$ into the two components of the force containing, respectively, only the symmetric matrices $\mathbf{F}_{\rm s}^{\rm Th}$, and the vector part $\mathbf{F}_{\rm a}^{\rm Th}$. Here we study $\mathbf{F}_{\rm a}^{\rm Th}$, and refer to Ref.~\cite{Reiche26} for the discussion of the first term. 

Using only symmetry properties of the relevant tensors and vector identities (see SM~\cite{SuppMat}), we rewrite the contribution to Eq.~\eqref{FCPNEqTH} stemming from skew-symmetric components of the tensors in Eq.~\eqref{decomposition} as 
\begin{align}
\label{Force_Decomposition}
\mathbf{F}^{\rm Th}_{\rm a}  
&= \int_{-\infty}^{\infty} d\omega\; \boldsymbol{\Omega}(\mathbf{R}_a,\omega, v)\times  \mathbf{J}_{\overline\chi}(\mathbf{R}_a, \omega,v)
\nonumber\\
&+\int_{-\infty}^{\infty} d\omega\; \left[ \boldsymbol{\Omega}(\mathbf{R}_a,\omega,v)\cdot\nabla_{\mathbf{R}_a} \right]\overline{\boldsymbol{\chi}}(\mathbf{R}_a, \omega,v).
\end{align}
In the equation above $\Omega$ is associated with the net rotation of the dipole vector operator $\hat{\mathbf{d}}(t)$ and we have introduced the net electromagnetic spin density
\begin{equation}\label{spin_density}
\overline{\boldsymbol{\chi}}(\mathbf{R}_a, \omega, v)=2\int_{-\infty}^{\infty}\frac{d q}{2\pi}\theta(-\omega_q^+)\boldsymbol{\chi}(q,\mathbf{R}_a,\omega_q^+)~,
\end{equation}
and the associated current density $ \mathbf{J}_{\overline\chi}=\nabla_{\mathbf{R}_a}\times\overline{\boldsymbol{\chi}}$. Equation~\eqref{Force_Decomposition} shows that the nonequilibrium Casimir-Polder force contains two contributions: a Magnus-like term and a Stern-Gerlach-like spin–gradient term~\cite{MolecularBeams21}. 

We focus on the Magnus-like force, which induces a drift orthogonal to the particle's direction of motion and is nonzero as long as neither $\boldsymbol{\Omega}$ nor $\overline{\boldsymbol{\chi}}$ vanish~\cite{Reiche20a}. Due to the symmetry of the system, both vectors are orthogonal to the direction of motion of the particle~\cite{Intravaia26}. The spin-gradient term, given by the directional derivative $\boldsymbol{\Omega}(\omega,v)\cdot\nabla_{\mathbf{R}_a}$, probes spatial variations of the electromagnetic spin density along the direction of the dipole's rotation, generating a force analogous to a Stern–Gerlach effect~\footnote{Classically, this corresponds to a force generated by shear, stratification, or other inhomogeneities in the fluid aligned with the object's spin direction. This contribution tends to align $\boldsymbol{\Omega}$ and $\overline{\boldsymbol{\chi}}$ and pulls the particle toward regions where the density of spin field along the direction of $\boldsymbol{\Omega}$ is strongest (if $\boldsymbol{\Omega}$ is prevalently aligned with $\overline{\boldsymbol{\chi}}$) or weakest (if it is prevalently anti-aligned).}. Unlike the Magnus-like term, it vanishes for geometries that are translationally invariant and mirror symmetric with respect to a plane containing the trajectory, such as a particle moving above a flat surface~\footnote{For these configurations, symmetry confines $\boldsymbol{\Omega}$ along a direction orthogonal to the symmetry plane~\cite{Intravaia26}, while the gradient is nonzero along a direction parallel to the plane, resulting in the vanishing of the directional derivative.}.

To understand the origin of these contributions, we examine the electromagnetic properties of the near-field region close to an interface with a material body. For an evanescent surface wave, $\bm{\chi}(q, \mathbf{R}_{a}, \omega)$ is orthogonal to the direction of propagation and, due to the spin-momentum locking of light~\cite{Bliokh15,Lodahl17}, its sign is tied to that of the wavevector. As a consequence, $\bm{\chi}(q, \mathbf{R}_{a}, \omega)$ is an odd function of $q$, and its integral over $q$ vanishes at equilibrium. In the reference frame of the moving particle, the Doppler shift breaks this balance, leading to a net spin density defined in Eq.~\eqref{spin_density}. When this spin density is not uniform, it generates a spin current density, $ \mathbf{J}_{\overline\chi}$, directed along the particle's trajectory and flowing around it. The net flow of this spin field acts as the relative velocity of the fluid with respect to the moving object. Unlike a classical fluid, however, this flow is provided by the quantum fluctuations of the electromagnetic field. 

The origin of $\boldsymbol\Omega$ follows from the interplay between the spin-momentum locking of light and the anomalous Doppler-effect~\cite{Nezlin76,Intravaia19a}. A neutral polarizable particle moving in the near field parallel to a material body interacts with the electromagnetic fluctuations supported by the interface. The combination of these two effects leads to asymmetric absorption and emission of linear and angular momentum by the particle~\cite{Intravaia19a}. As a result, the particle acquires a net angular momentum, captured by $\boldsymbol{\Omega}$. This phenomenon can be interpreted as part of the energy supplied by the external drive to the system being converted into ``rotational'' kinetic energy. Since the interaction strength depends on the distance from the surface, $\boldsymbol{\Omega}$ also depends on $\mathbf{R}_a$. 

To illustrate the system's dynamics and to highlight the analogy with the classical rotational effect defining the Magnus force, we model the particle's dipole as a three-dimensional quantum harmonic oscillator with transition frequency $\omega_a$. In this case, the nonequilibrium power spectrum takes the form~\cite{Reiche20a}
\begin{subequations}
\label{SandAlpha}
\begin{equation}
\underline{S}(\omega,v)
=
\frac{\hbar}{\pi}
\int \frac{dq}{2\pi}
\theta(\omega^+_{q})
 \underline{\alpha}(\omega,v)
\underline{G}_{\Im}(q,\mathbf{R}_a,\omega^+_{q})
\underline{\alpha}^\dagger(\omega,v),
\label{Sneq}
\end{equation}
where $\underline{\alpha}_{\Im}=[\underline{\alpha}-\underline{\alpha}^{\dagger}]/2\imath$ and 
\begin{equation}
\label{susceptibility}
    \underline{\alpha}(\omega,v) = \frac{\alpha_B(\omega)}{1-\alpha_B(\omega)\int \frac{dq}{2\pi} \underline{G}(q, \mathbf{R}_a,\omega_{q}^+)}
\end{equation}
\end{subequations}
is the resulting dressed velocity-dependent polarizability tensor~\cite{Intravaia16}. In Eqs.~\eqref{SandAlpha} $\alpha_B(\omega)=\alpha_0 \omega_a^2/(\omega_a^2-[\omega+\imath 0^{+}]^2)$ and $\alpha_0$ are, respectively, the (causal) bare and the static oscillator's polarizabilities, $\theta(\omega)$ is the Heaviside step-function and $\underline{G}_{\Im}$, defined in analogy to $\underline{\alpha}_{\Im}$, is proportional to the skew-hermitian part of the Fourier transformed Green tensor. 
The nonequilibrium fluctuation theorem expressed in Eqs.~\eqref{SandAlpha} explicitly includes atomic rotational degrees of freedom~\cite{Intravaia19a}. These expressions enable us to evaluate the force in Eq.~\eqref{Force_Decomposition}. Similar to previous work~\cite{Intravaia19a}, we consider a $\ce{^{87}Rb}$ atom moving along the positive $x$-direction at a constant height $z_a>0$ within the near field of a planar interface with a bulk ($z_a\le 0$) composed of a local material. For this geometry, the spin-gradient term in Eq.~\eqref{Force_Decomposition} identically vanishes, and the electromagnetic Green tensor is known analytically~\cite{Intravaia16,Wylie85}. In the near field limit, which dominates the interaction, we have 
\begin{equation}
\label{GT}
\begin{split}
&\underline{G}(q, z_a, \omega)
\;\sim\;
\frac{r(\omega)}{2 \varepsilon_0} \int\frac{dp}{2\pi}\;\underline{\Pi}\;k
\;e^{-2 k z_a},\\
& {\rm where} \quad 
\underline{\Pi}=\begin{pmatrix}
\frac{q^2}{k^2} & \frac{pq}{k^2} & -\rm{i}\frac{q}{k} \\
\frac{pq}{k^2} & \frac{p^2}{k^2} & -\rm{i}\frac{p}{k} \\
\rm{i}\frac{q}{k} & \rm{i}\frac{p}{k} & 1
\end{pmatrix} \xrightarrow{} \begin{pmatrix}
\frac{q^2}{k^2} & 0 & -\rm{i}\frac{q}{k} \\
0 & \frac{p^2}{k^2} & 0 \\
\rm{i}\frac{q}{k} & 0 & 1
\end{pmatrix}.
\end{split}
\end{equation}
Here $k=\sqrt{q^2+p^2}$, $\varepsilon_0$ is the vacuum permittivity, $r(\omega)$ the surface’s p-polarized reflection coefficient and $\underline{\Pi}$ the corresponding dyadic tensor (the contribution of s-polarization is negligible). From $\underline{\Pi}$ we have removed the terms that do not contribute, due to symmetry \footnote{Here we consider only the scattering part of the Green tensor, which encodes the reflection from the planar surface; the vacuum contribution represents a constant energy shift and does not play a role in the nonequilibrium force studied in this work.}.
In the atom's rest frame, surface waves predominantly generate a spin density oriented along the negative y-direction, which in turn induces a current $\mathbf{J}_{\overline\chi}$ along the negative $x$-direction. Within the hydrodynamic analogy, this configuration can be interpreted as a quantum fluid flowing against the atom. In this sense, the electromagnetic vacuum effectively behaves as a fluid and exerts a viscous resistance on the moving particle. Moreover, as the particle moves, it mainly absorbs surface excitations with negative spin, giving rise to a clockwise rotation of the dipole along the $y$-axis (see Fig.~\ref{fig:Magnus_force})~\cite{Intravaia19a}. This phenomenon implies that, just as $\overline{\boldsymbol{\chi}}$, also $\boldsymbol\Omega$ is oriented along the negative $y$-axis (see SM~\cite{SuppMat}). In direct analogy with the classical Magnus effect, the interplay between this induced rotation and the translational motion generates a transverse force. As expressed in Eq.~\eqref{Force_Decomposition}, this contribution naturally takes the form of a cross product, it acts perpendicular to the velocity and lifts the atom away from the surface.

\begin{figure}
   \includegraphics[width=1.\linewidth]{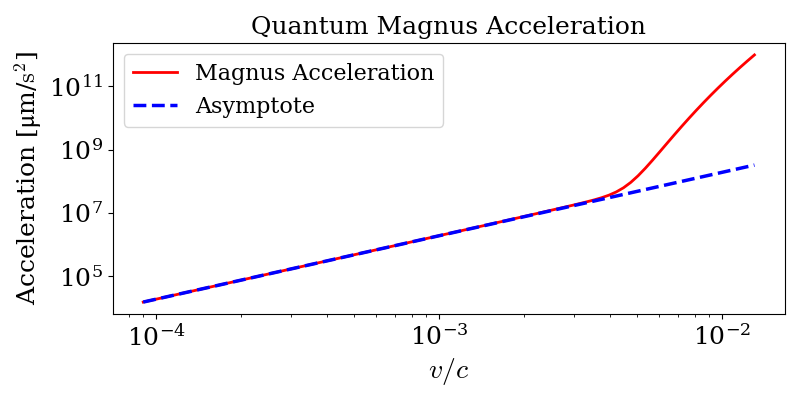}
  \vspace{-0.6cm}
  \caption{Acceleration $a_{\rm a} = F^{\rm Th}_{\rm a}/m_\mathrm{Rb}$ for a $\ce{^{87}Rb}$ atom ($\alpha_0 = 4\pi\varepsilon_0 \times 47.28~\text{\AA}^3$, $\omega_a = 1.6~\mathrm{eV}$, $m_\mathrm{Rb} = 86.9~\mathrm{u}$~\cite{Intravaia16}) as a function of its velocity corresponding to the Magnus-like force in Eq.~\eqref{Force_Decomposition}. The atom moves in vacuum within the near field ($z_a = 5~\mathrm{nm}$) of a planar vacuum-gold interface described by a Drude permittivity $\varepsilon(\omega) = 1 - \omega_p^2/[\omega(\omega + \imath \gamma)/$ ($\omega_p = 9~\mathrm{eV},~\gamma = 35~\mathrm{meV}$~\cite{Intravaia19a}, $\rho = \gamma/[\varepsilon_0 \omega_p^2] = 3.21\times 10^{-8}~\Omega\,\mathrm{m}$). At low velocity, the acceleration (red solid curve) scales as $\propto v^2/z_a^9$, in agreement with the corresponding low-velocity asymptotic expression, Eq.~\eqref{asymptote} (blue dotted line). 
    \label{fig:Magnus_acceleration}
  }
  \vspace{-0.5cm}
\end{figure}
\begin{figure}
   \includegraphics[width=1.\linewidth]{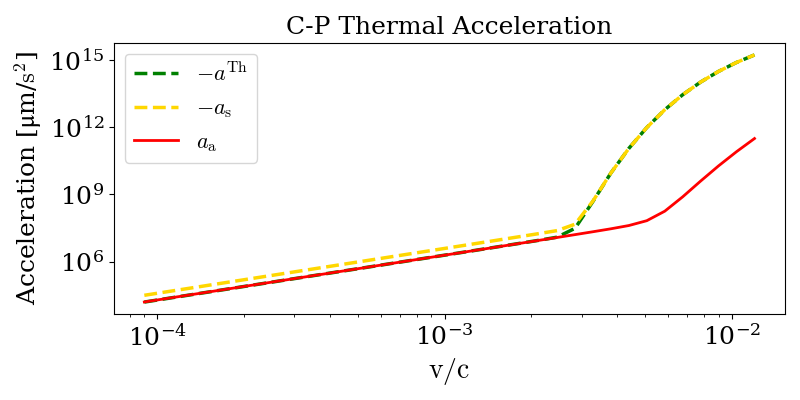}
  \vspace{-0.6cm}
  \caption{Thermal component of the nonequilibrium Casimir-Polder acceleration $a^{\rm Th} = F^{\rm Th}/m_\mathrm{Rb}$ for a $\ce{^{87}Rb}$ atom moving in vacuum within the near field of a planar metallic interface (see Fig.~\ref{fig:Magnus_acceleration}). The different components $a^{\rm Th} = a_{\rm a} +a_{\rm s}$, 
 are shown as a function of the particle's velocity. 
    \label{fig:Thermal_acceleration}
  }
\end{figure}

In Fig.~\ref{fig:Magnus_acceleration} we depict the behavior of the antisymmetric component of the nonequilibrium thermal acceleration in Eq.~\eqref{FCPNEqTH} as a function of the particle's velocity. In accordance with the previous analysis, this results in a force directed along the $z$-direction that acts as a repulsive force with respect to the surface, providing a clear signature of a Magnus-like effect. 
The curve shows two different regimes: for sufficiently low velocities the force features a power-law behavior; In contrast, for relatively high velocities, the force grows exponentially. For a quantitative analysis, we focus on the low-velocity region. In this regime, an expansion in the particle velocity is justified provided that the frequency correction induced by the Doppler shift is sufficiently small~\cite{Reiche22} (see SM~\cite{SuppMat}). In the near field regime, the dominant contribution originates from $q\sim 1/z_a$, leading to the condition that $v/z_a$ has to be smaller than the characteristic frequencies of the system, such as atomic transition frequencies and material resonances~\cite{Reiche22}. Within this velocity range, the force in Eq.~\eqref{Force_Decomposition} is normal to the surface plane, and, to second order in $\alpha_0$, it can be written as (see SM~\cite{SuppMat})
\begin{equation}
\label{asymptote}
   F^{\rm Th}_{\rm a} \sim \frac{9\hbar \alpha_0^2 \rho}{\pi^3  \varepsilon_0}\frac{v^2}{(2z_a)^9}.
\end{equation}
The parameter $\rho\equiv\partial_{\omega}\mathrm{Im}[r(\omega)/(2\varepsilon_{0})]_{\vert\omega=0}$ is related to the density of state of the electromagnetic field at low frequency~\cite{SuppMat}.
Specifically, for a metallic material, $\rho$ can be identified as its resistivity. Equation~\eqref{asymptote} demonstrates that, in agreement with the intrinsic nature of the interaction mentioned above, the Magnus-like term scales quadratically with the velocity of the particle. This result implies that reversing the particle's direction of motion does not change the sign of the drift, which continues to repel the particle from the surface. In the present nonequilibrium setting, indeed, the anomalous Doppler effect and spin-momentum locking of light generate a net atomic angular momentum scaling as $\propto v$ and whose orientation is determined by the direction of motion~\cite{Intravaia19a}. When the particle reverses its trajectory, the Doppler shift reverses, thereby reversing both $\boldsymbol{\Omega}$ and $\mathbf{J}_{\overline{\chi}}$. Their cross product, and therefore the direction of the Magnus-like force remains unchanged. As a consequence, the force must be an even function of the particle's velocity.
In contrast, if the angular momentum were independent of the translation direction, as for an externally driven rotation or due to an intrinsic spin, the Magnus-like term would reverse direction, inducing an attractive contribution. The force further decays with the atom–surface separation as $\propto z_{a}^{-9}$, reflecting the near-field character of the interaction, dominated by evanescent electromagnetic modes. 
Finally, in Fig.~\ref{fig:Thermal_acceleration} we compare the behavior of the symmetric and antisymmetric components of the acceleration arising from Eq.~\eqref{FCPNEqTH} as a function of the particle's velocity. The symmetric contribution (yellow dotted curve) yields an attractive force towards the surface, which is consistent with the Casimir-Polder interaction. 

In summary we have shown that a particle moving in a quantum vacuum modified by the presence of macroscopic bodies, experiences a Magnus-like transverse force. 
Specifically, we demonstrated that the force arises from the interplay between the spin of the electromagnetic field and the particle’s rotational degrees of freedom, appearing as a distinct contribution to the nonequilibrium Casimir–Polder interaction.
It is worth noting that the underlying mechanism shares analogies with the spin Hall effect in condensed matter systems~\cite{Dyakonov71,Dyakonov71a,Hirsch99,SpinPhysics17,Chumak15,Litzius17}, while being more directly connected to the spin Hall effect of light~\cite{Bliokh15}. In this sense, the force reflects the nonequilibrium spin-orbit dynamics induced by the particle's motion and its interaction with the material-modified electromagnetic quantum fluctuations. However, since the particle's spin is induced and its sign uniquely determined by the direction of motion, reversing the velocity of the particle leaves unchanged the direction of the Magnus-like force. This connection also opens the way to study this force as a Lorentz force, produced by the Berry curvature acting in momentum space~\cite{Bliokh15}.

Our work goes beyond previous studies on motion-induced nonequilibrium atom-surface interactions~\cite{Shresta03,Dedkov21,Reiche26}. It is independent of specific assumptions about local thermal equilibrium or Markovianity, and applies to a broad class of particles and electromagnetic environments~\cite{Intravaia16}. We emphasize the quantum origin of the phenomenon described here, in which the rotation emerges from the interaction with the electromagnetic field under nonequilibrium driving conditions~\cite{Intravaia19a,Oelschlager22}. In previous works, instead, rotation is typically externally imposed through direct driving~\cite{Amaral25,Deop-Ruano23,Zhao12,Volokitin18,Dedkov12,Dedkov16,Dedkov17}. In contrast, in our case it arises from the interaction within the system and the parametric amplification of quantum fluctuations~\cite{Oelschlager22,Nation12}. As an example, we have considered an atom moving at a constant velocity above a planar metallic interface. We have shown that in this geometry the resulting Magnus-like force induces a lift-up in the particle's motion opposite to the equilibrium attractive Casimir-Polder force.

Detecting the Magnus-like force in relation to the other components of the Casimir-Polder force, particularly in the case of $\mathbf{F}^{\rm Ds}$, can be challenging due to the relatively weak nature of the effect. 
Nevertheless, several strategies can enhance its observability. For instance, increasing the particle's velocity strengthens the Magnus-like contribution, while previous studies have shown that $\mathbf{F}^{\rm Ds}$ tends to decrease in the same limit~\cite{Reiche26}. 
Atoms moving at high speeds can be realized using thermal beams~\cite{Grisenti99,Perreault05,Garcion21} or by neutralizing ion beams~\cite{Schuller07,Rousseau07,Kanitz25} accelerated using ion guns~\cite{Kanitz25}, linear~\cite{Wei12} or circular~\cite{Denker23} accelerators. Furthermore, benefiting from the analogy with the spin Hall effect, other approaches can rely on the vectorial nature of the interaction, in particular its dependence on the system's geometry and material properties~\cite{Joseph24,Kipp21,Manchon19,Tang24}. 
Engineering these properties can lead to an additional increase in the interaction strength, which also helps to isolate the distinct signature of the Magnus-like force. For example, enhancing the dissipative response of the material can effectively increase the coefficient $\rho$ and therefore the force in Eq.~\eqref{asymptote}.
Although an accurate detection of atom-surface interaction can be challenging, sensitive experiments with ultracold atoms~\cite{Bender14}, spectroscopy methods with nanoscopic cells~\cite{Whittaker14,Carvalho18} and atomic diffraction patterns~\cite{Lonij09, Lecoffre25}, have already demonstrated the feasibility of observing Casimir–Polder forces with high precision. These developments suggest that carefully designed setups, in principle, allow for the detection of the Magnus-like transverse drift predicted here.

\paragraph{Acknowledgements}~-- This project is financially supported by BERLIN QUANTUM. An initiative endowed by the Innovation Promotion Fund of the city of Berlin. F.I. acknowledges and sincerely thank Diego A. R. Dalvit for
motivating questions and valuable discussions.
\bibliographystyle{prstytitlenew}
\bibliography{bibliomia}

\cleardoublepage






\newpage
\makeatletter

\renewcommand\section{\@startsection {section}{1}{\z@}%
                                   {-3.5ex \@plus -1ex \@minus -.2ex}%
                                   {2.3ex \@plus.2ex}%
                                   {\center\normalfont\large\bfseries}}

\makeatother



\setcounter{page}{1}
\setcounter{figure}{0}
\setcounter{equation}{0}
\renewcommand{\theequation}{S\arabic{equation}}
\renewcommand{\figurename}{\textbf{Supplementary Figure}}
\renewcommand{\thefigure}{{\bf S\arabic{figure}}}
\renewcommand{\thefigure}{{\bf S\arabic{figure}}}

\begin{center}
\section{Supplemental Material}
\textit{\large\bf Nonequilibrium Casimir-Polder Force: Magnus-Like Effect\vspace{0.2cm}}

\noindent M. V. Gurrieri$^{1}$, K. Busch$^{1,2}$ \\
and F. Intravaia$^{1}$.
\end{center}
\begin{footnotesize}
\begin{enumerate}
\item[$^{1}$]
Humboldt-Universit\"at zu Berlin, Institut f\"ur Physik, AG Theoretische Optik and Photonik, 12489 Berlin, Germany
\item[$^{2}$]
Max-Born-Institut, 12489 Berlin, Germany
\end{enumerate}
\end{footnotesize}

In this supplementary material, we provide further details to support the conclusions in the main text, discussing some of the expressions presented in the manuscript.

\subsection{The thermal component of Casimir-Polder force}
\label{force_derivation}
The derivation of the nonequilibrium thermal component of the Casimir-Polder force, Eq.~(2) of the main text, is shortly reviewed here. A more detailed derivation of the full nonequilibrium force, Eq.~(1), can be found in the Supplemental Material of Ref.~\cite{Reiche26}.
\begin{widetext}
The starting point is the formulation of the Lorentz force $\mathbf{F}(t)=\lim_{\mathbf{r}\rightarrow\mathbf{r}_a(t)}\sum_{i} \left\langle\hat{d}_{i}(t)\nabla_{\mathbf{R}}\hat{E}_{i}(\mathbf{r},t)\right\rangle$, where $\mathbf{r}_a(t)$ is the particle's trajectory, $\hat{d}_{i}$ and $\hat{E}_{i}$ are the $i$-component of electric dipole vector operator and of the total electric field operator, respectively. As shown in Refs.~\cite{Intravaia16,Reiche26}, in the stationary limit, the total force can be expressed as follows:
\begin{align}
\label{Force_Casimir_Polder}
\mathbf{F}
				&=\int\limits_{0}^{\infty}\mathrm{d}\omega\int \frac{\mathrm{d}q}{2\pi}\;
						\frac{\hbar}{2\pi}\mathrm{Im}\mathrm{Tr}\left[
						\underline{\alpha}_{v}(\omega_{q}^{-}) \nabla_{\mathbf{R}_{a}}\underline{G}(q, \mathbf{R}_{a}, \omega)\right]
					\\
				&+\frac{1}{2}\int \frac{\mathrm{d}q}{2\pi}\int \mathrm{d}\omega\,\mathrm{Tr}\left[
\left(\frac{\hbar}{\pi}\left\{ \theta(\omega_{q}^{-})-\theta(\omega)\right\}
\underline{\alpha}_{\Im}(\omega_{q}^{-})+\underline{J}(\omega_{q}^{-},v) \right)\nabla_{\mathbf{R}_{a}}\underline{G}_{\Re}(q, \mathbf{R}_{a}, \omega)\right]+\mathbf{F}^{\rm As}
\nonumber
\end{align}
\end{widetext}
where $\omega_{q}^{\pm}=\omega \pm qv$ is the Doppler-shifted frequency. 
The first and last terms in Eq.~\eqref{Force_Casimir_Polder} correspond, respectively, to $\mathbf{F}^{\rm Ds}$ and to the geometric contributions to the Casimir-Polder force. We focus on the second term. Using the Kramers-Kronig relations for the atomic polarizability $\underline{\alpha}$ and the Green tensor $\underline{G}$~\cite{Jackson75,Landau80}, together with the Hermitian properties of the tensors $\underline{S}$ and $\underline{J}$, and noting that $\underline{J}$ satisfies the symmetry relation $\underline{J}(\omega,v)=\underline{J}^{*}(-\omega,v)$~\cite{Intravaia26}, in analogy with $\underline{\alpha}$ and $\underline{G}$, the second term on the right-hand side of Eq.~\eqref{Force_Casimir_Polder} can, after some algebra, be rewritten in the form of Eq.~(2) of the main text. This is equivalent to
\begin{align}
\label{FcpThermal_sym_asym}
   \mathbf{F}^{\rm Th} = \int d \omega 
      \int\frac{d q}{2 \pi} &\; \theta(-\omega_q^+)
      \nonumber\\
      &\operatorname{Tr}\left[\underline{S}(\omega)\nabla_{\mathbf{R}_a} \underline{G}_{\Re}(q, \mathbf{R}_a \omega_q^+) \right].
\end{align}

\subsection{Vectorial form of $\mathbf{F}_{\rm a}^{\rm Th}$}
\label{model} 
As mentioned in the main text, $\underline{S}(\omega)$ is Hermitian and thus it can be written as $\underline{S}(\omega) = \underline{K}(\omega) + \mathbf{\Omega}(\omega)\cdot \underline{\bm{L}}$, where $\underline{K}(\omega)$ is a real symmetric matrix, $\mathbf{\Omega}(\omega)$ is a real vector, and $\left[ L_k \right]_{ij} =- i \epsilon_{kij}$ is the skew-symmetric matrix generator of the Lie algebra $\bm{\mathfrak{so}(3)}$, with $\epsilon_{kij}$ the Levi-Civita symbol \cite{Intravaia26}. Given that $\underline{G}_{\Re}(q, \mathbf{R}_a, \omega_q^+)$ is also Hermitian it can be written as the sum of a real-symmetric and skew-symmetric matrix, respectively $\underline{\Sigma}(q, \mathbf{R}_a, \omega_q^+)$ and $\bm{\chi}(q, \mathbf{R}_a, \omega_q^+)\cdot \underline{\bm{L}}$. Since the trace of a symmetric and antisymmetric matrix is zero, $\mathbf{F}^{\rm Th}$ decomposes into two terms, one of which is
\begin{align}
\label{FcpThermal_asym}
   \mathbf{F}^{\rm Th}_{a} = \int d \omega 
      \int\frac{d q}{2 \pi} &\; \theta(-\omega_q^+)
      \nonumber\\
      &\operatorname{Tr}\left[\mathbf{\Omega}(\omega)\cdot \underline{\bm{L}} \nabla_{\mathbf{R}_a} \big(\bm{\chi}(q, \mathbf{R}_a, \omega_{q}^+)\cdot \underline{\bm{L}}\big) \right].
\end{align}
Using vector identities, $\mathbf{F}^{\rm Th}_{a}$ can be expressed in the form given in Eq.~(5) of the main text.
All the involved functions depend only on the lateral coordinate $\mathbf{R}_a$, so the gradient operator can be replaced by its full three-dimensional form, i.e. $\nabla_{\mathbf{R}_a}\to \nabla$.
Moreover, since $\bm{\chi}$ and $\bm{\Omega}$ are the only quantities with functional dependencies, we omit their arguments whenever there is no ambiguity. We can then rewrite the gradient as follows
\begin{equation}
    \nabla\big(\bm{\chi}\cdot \underline{\bm{L}}\big)=\big(\nabla\bm{\chi}^{\sf T}\big)\cdot \underline{\bm{L}}.
\end{equation}
Given that $\operatorname{Tr}\left[\underline{L}_i\underline{L}_j\right]=2\delta_{ij}$, we then obtain
\begin{equation}
     \operatorname{Tr}\left[\mathbf{\Omega}\cdot \underline{\bm{L}} \big(\nabla\bm{\chi}^{\sf T}\big)\cdot \underline{\bm{L}} \right]
\\
= 2\big(\nabla \bm{\chi}^{\sf T}\big)\cdot \bm{\Omega},
\end{equation}
where $\nabla\bm{\chi}^{\sf T}$ is a $3\times 3$ dyadic. Using the identity $\mathbf{A}\times ( \nabla \times \mathbf{B}) = (\nabla\mathbf{B}^{\sf T})\cdot \mathbf{A}-(\nabla\mathbf{B}^{\sf T})^{\sf T} \cdot\mathbf{A}$, it follows that
\begin{equation}\label{splitting_force}
\begin{split}
    \operatorname{Tr}\left[\mathbf{\Omega}\cdot \underline{\bm{L}} \nabla\big(\bm{\chi}\cdot \underline{\bm{L}}\big) \right]&= \boldsymbol{\Omega}\times \left( 2\, \nabla \times \boldsymbol{\chi} \right) 
     +\left(2\, \nabla \,\boldsymbol{\chi}^{\sf T} \right)^{\sf T} \cdot\boldsymbol{\Omega}.
    \end{split}
\end{equation}
When inserted into Eq.~\eqref{FcpThermal_asym}, the resulting expression is equivalent to Eq.~(5) of the main text.
The same result can also be obtained throughout a similar derivation using explicit index notation.
Notice that in the previous derivation we did not rely on any particular model for the particle’s internal dynamics or 
on the specific material properties and geometry of its electromagnet surroundings.

\subsection{Asymptotic analysis}\label{Asymptote_derivation}

\begin{figure}
   \includegraphics[width=1.\linewidth]{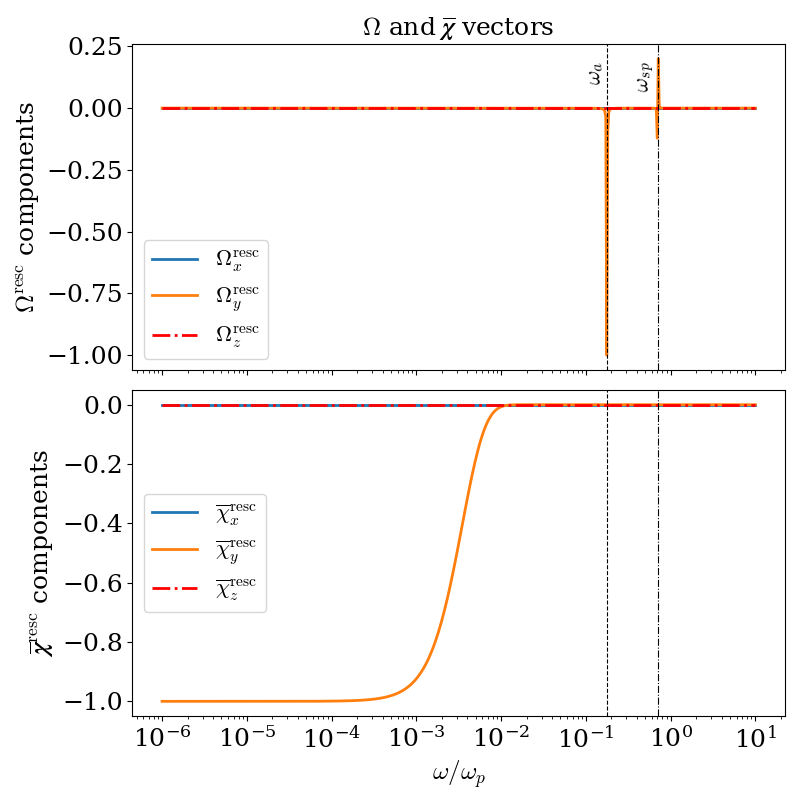}
  \vspace{-0.6cm}
  \caption{Behavior of the frequency-dependent components of $\overline{\boldsymbol{\chi}}(\omega)$ and $\boldsymbol{\Omega}(\omega)$ vectors for a particle moving at constant velocity above a planar gold surface. The characteristic atomic frequency $\omega_a$ corresponds to a rubidium atom (see main text), while $\omega_{sp}$ denotes the surface plasmon frequency of gold. Only the $y$-components of both vectors are non-zero. $\Omega_y$ and $\overline{\chi}_y$ are normalized by the absolute value of their minimum, i.e. $\Omega^{\mathrm{resc}}_y(\omega) = \Omega_y(\omega)/|\min(\Omega_y)|$ and $\overline{\chi}^{\mathrm{resc}}_y(\omega) = \overline{\chi}_y(\omega)/|\min(\overline{\chi}_y)|$, such that all plotted quantities are dimensionless.}
    \label{fig:real_vectors}
\end{figure}

In this section, we derive the asymptotic expansion of the Magnus-like force given in Eq.~(8) of the main text, which is used to validate our numerical results in Fig.~2.
Specifically, we consider the configuration in which the particle moves at a constant height within the near field of a vacuum–material interface formed by a local medium. In addition, following previous works, the internal dynamics of the particle are modeled as a three-dimensional harmonic oscillator~\cite{Intravaia19a}. 
Within this description, to second order in $\alpha_0$, the particle's power spectrum in Eq.~(6a) of the main text can be written as~\cite{Reiche20a}
\begin{align}
\label{spectrum}
\underline{S}(\omega, v) 
&\sim 
|\alpha_B(\omega)|^2\frac{\hbar}{\pi} 
\int \frac{dq}{2\pi} 
\theta(\omega_{q}^{+})   
\underline{G}_{\Im}(q, \mathbf{R}_a,\omega_{q}^{+}).
\end{align}
As also pointed out there, for a planar surface the near-field limit of the Green tensor simplifies considerably [see Eq.~(7) of the main text]. For this geometry we are 
going to consider that the interface lies at $z=0$ and the particle moves in vacuum at $z_{a}>0$ along the positive $x$-direction ($v>0$). For this system the vectors $\mathbf{\Omega}$ and $\overline{\boldsymbol{\chi}}$ are oriented along the $y$-direction while the spin current $\mathbf{J}_{\overline\chi}$ is along the $x$-direction, see Fig.~\ref{fig:real_vectors}. 
We can write Eq.~(5) as
\begin{equation}\label{Magnus_vector_product}
\mathbf{F}^{\rm Th}_{a}
=
\int d\omega 
\; \Omega(z_a,\omega) \bm{\bar y}\times\; \left[- \partial_{z_a} \overline{\chi}(z_a,\omega) \bm{\bar x}\right]
,
\end{equation}
where we introduced the unit vectors $\bm{\bar x}$,  $\bm{\bar y}$ and $ \bm{\bar z}$ parallel to
the directions of the reference frame.
To second order in $\alpha_0$ the magnitudes of the vectors are 
\begin{subequations}
\begin{equation}
\Omega(z_a,\omega)
\sim -\frac{\hbar}{\pi}
|\alpha_B(\omega)|^2
\int
\frac{dq}{2\pi}\,
\frac{q|q|K_1(2|q|z_a)}{2\pi\varepsilon_0}
r_{I}(\omega_{q}^{+})
\theta(\omega_{q}^{+}),
\end{equation}
\begin{equation}
\overline{\chi}(z_a, \omega)
\sim -2
\int
\frac{d\tilde{q}}{2\pi}\,
\frac{\tilde{q}|\tilde{q}|K_1(2|\tilde{q}|z_a)}{2\pi\varepsilon_0}
r_{R}(\omega_{\tilde{q}}^{+})
\theta(-\omega_{\tilde{q}}^{+}),
\end{equation}
\end{subequations}
where  $\varepsilon_0$ is the vacuum permittivity, and we used that for $k=\sqrt{p^{2}+q^{2}}$
\begin{equation}
\int\frac{d p}{2\pi}\;q\;
e^{-2kz_a} 
= \frac{2q |q|}{2\pi} 
K_1(2|q|z_a),
\end{equation} 
where $K_1(x)$ is the first order modified Bessel function of the second kind. In the previous expressions, $r_{\rm R}(\omega)$ and $r_{\rm I}(\omega)$ represent the real and imaginary components of the near-field limit of the planar interface’s p-polarized reflection coefficient, respectively (the s-polarized contribution can be neglected in the near-field limit). 

An expansion in the particle velocity is justified when the Doppler shift, $|qv|$, is smaller than all the system's characteristic frequencies $\omega_{\mathrm{eff}}$. Indeed, in the expression for the force in Eq.~\eqref{Magnus_vector_product}, the Heaviside theta functions limit the integration over the frequency to the interval $-qv<\omega<-\tilde{q}v$. Due to the Bessel function, the dominant contribution to the integrals arises from $|q| \sim 1/z_a$. It follows that, within the low-velocity of our nonrelativistic approximation, $|\omega|<v/z_a \ll \omega_{\mathrm{eff}}$. This condition defines the velocity range over which the low-velocity expansion is valid and constrains the relevant frequencies to a relatively small range around zero.
The same smallness constraint is also valid for $\omega_{q}^{\pm}$ which allows us to consider the following expansions
\begin{equation}
\label{material}
r_{R}(\omega_{\tilde q}^{+}) \sim r(0), \quad r_{I}(\omega_{q}^{+}) \sim r'_{I}(0)\omega_{q}^{+},\quad
|\alpha_B(\omega)|^2 \sim \alpha_0^2,
\end{equation}
where the prime indicates the derivative with respect to the argument. Notice that these expansions are valid, in general, for a generic local ohmic material.
Substituting the previous low-frequency expansion back into Eqs.~\eqref{Magnus_vector_product}, the integral in $\omega$ can be performed as follows 
\begin{equation}
\begin{split}
\int d\omega \;\omega_{q}^{+}&\theta(\omega_{q}^{+})\theta(-\omega_{\tilde{q}}^{+}) 
\stackrel{\omega_{q}^{+}=\tilde{\omega}}{=}
\int d\tilde{\omega} \;\tilde{\omega}\;\theta(\tilde{\omega})\theta(-[\tilde{\omega}-(q-\tilde{q})v])
\\
&=\theta(q-\tilde{q})\int\limits_{0}^{(q-\tilde{q})v} d\tilde{\omega} \;\tilde{\omega}=\theta(q-\tilde{q})\frac{(q-\tilde{q})^{2}v^{2}}{2}.
\end{split}
\end{equation} 

Collecting all the expressions we get
\begin{widetext}
\begin{equation}
\label{Magnus_last_integral}
\begin{split}
\mathbf{F}^{\rm Th}_{a} \sim&- \left[
 -2\frac{\hbar}{\pi}
\alpha_{0}^2 r'_{I}(0)r_{R}(0)
\int
\frac{dq}{2\pi}\,
\frac{q|q|K_1(2|q|z_a)}{2\pi\varepsilon_0}
\int
\frac{d\tilde{q}}{2\pi}\,
\frac{\tilde{q}|\tilde{q}|\partial_{z_{a}}K_1(2|\tilde{q}|z_a)}{2\pi\varepsilon_0}
\theta(q-\tilde{q})\frac{(q-\tilde{q})^{2}v^{2}}{2}\right] \bm{\bar z},
\\
=-&
 \frac{\hbar}{\pi}
\frac{\alpha_{0}^2}{\varepsilon_0^{2}} \frac{4 r'_{I}(0)r_{R}(0)v^{2}}{(2\pi)^{4}}
\int\limits_{0}^{\infty}
dq\,
q^{3}K_1(2qz_a)
\int\limits_{0}^{\infty}
d\tilde{q}\,
\tilde{q}^{3}\partial_{z_{a}}K_1(2\tilde{q}z_a)
 \bm{\bar z}=  \hbar\frac{\alpha_0^2 }{\epsilon_0^2}\frac{9}{2\pi^3}
r'_{\rm I}(0)r(0)\frac{v^2}
{(2z_a)^9}
\bm{\bar z}.
\end{split}
\end{equation}
\end{widetext}
For an ohmic metallic interface, the reflection is saturated so that $r_{R}(0)=1$. In addition, we can identify $\rho=r'_{\rm I}(0)/(2\varepsilon_{0})$ with the resistivity of the conductor so that we recover Eq.~(8) in the main text.

For clarity, in Fig.~\ref{fig:real_vectors}, we plot the behavior of the components of the vectors $\overline{\boldsymbol{\chi}}$ and $\boldsymbol{\Omega}$ as a function of frequency for a particle moving at constant speed above a planar surface. As reported in the main text, in the rest frame of the atom, surface waves appear predominantly with a spin oriented along the negative y-direction, so that the net spin density $\overline{\boldsymbol{\chi}}$ is negative. The particle mainly absorbs surface excitations with negative spin, leading to a clockwise ``rotation" of the dipole about the y-axis (see Fig.~1 in the main text). 
\newpage


\end{document}